\documentclass[aps,prc,preprint,doublespaced,superscriptaddress,nofootinbib,
               showpacs,showkeys]{revtex4}

\usepackage{graphicx}

\begin{document}


\title{
Measurements of total and partial charge-changing cross sections
for 200-400~MeV/nucleon $^{12}$C in water and polycarbonate 
}

%
%

\newcommand{\authorjst}
{CREST, Japan Science and Technology Agency, Kawaguchi 332-0012, Japan}

\newcommand{\authorkek}
{High Energy Accelerator Research Organization (KEK), Tsukuba 305-0801, Japan}

\newcommand{\authoraue}
{Aichi University of Education, Kariya 448-8542, Japan}

\newcommand{\authorchalmers}
{Chalmers University of Technology SE-412 96 Gothenburg, Sweden,\\
Roanoke College, Salem, VA 24153, USA}

\newcommand{\authorgunma}{Gunma University, Maebashi 371-8510, Japan}

\newcommand{\authorjaxa}
{Japan Aerospace Exploration Agency (JAXA), Sagamihara 229-8510, Japan}

\newcommand{\authorkobe}
{Kobe University, Kobe 657-8501, Japan}

\newcommand{\authornagoya}{Nagoya University, Nagoya 464-8602, Japan}

\newcommand{\authornaruto}{Naruto University of Education, Naruto 772-8502, Japan}

\newcommand{\authornirs}
{National Institute of Radiological Sciences (NIRS), Chiba 263-8555, Japan}

\newcommand{\authorslac}
{Stanford Linear Accelerator Center (SLAC), Stanford, CA 94309, USA}

\newcommand{\authortoho}{Toho University, Funabashi 274-8510, Japan}

\author{T.~Toshito}
\affiliation{\authorjst}
\affiliation{\authorkek}

\author{K.~Kodama}
\affiliation{\authoraue}

\author{L.~Sihver}
\affiliation{\authorchalmers}

\author{K.~Yusa}
\affiliation{\authorgunma}

\author{M.~Ozaki}
\affiliation{\authorjaxa}

\author{K.~Amako}
\affiliation{\authorkek}

\author{S.~Kameoka}
\affiliation{\authorkek}

\author{K.~Murakami}
\affiliation{\authorkek}

\author{T.~Sasaki}
\affiliation{\authorkek}

\author{S.~Aoki}
\affiliation{\authorkobe}

\author{T.~Ban}
\affiliation{\authornagoya}

\author{T.~Fukuda}
\affiliation{\authornagoya}

\author{M.~Komatsu}
\affiliation{\authornagoya}

\author{H.~Kubota}
\affiliation{\authornagoya}

\author{N.~Naganawa}
\affiliation{\authornagoya}

\author{T.~Nakamura}
\affiliation{\authornagoya}

\author{T.~Nakano}
\affiliation{\authornagoya}

\author{M.~Natsume}
\affiliation{\authornagoya}

\author{K.~Niwa}
\affiliation{\authornagoya}

\author{S.~Takahashi}
\affiliation{\authornagoya}

\author{J.~Yoshida}
\affiliation{\authornagoya}

\author{H.~Yoshida}
\affiliation{\authornaruto}

\author{M.~Kanazawa}
\affiliation{\authornirs}

\author{N.~Kanematsu}
\affiliation{\authornirs}

\author{M.~Komori}
\affiliation{\authornirs}

\author{S.~Sato}
\affiliation{\authornirs}

\author{M.~Asai}
\affiliation{\authorslac}

\author{T.~Koi}
\affiliation{\authorslac}

\author{C.~Fukushima}
\affiliation{\authortoho}

\author{S.~Ogawa}
\affiliation{\authortoho}

\author{M.~Shibasaki}
\affiliation{\authortoho}

\author{H.~Shibuya}
\affiliation{\authortoho}


%
%

\begin{abstract}
We have studied charged nuclear fragments produced by 200 - 400 MeV/nucleon carbon ions,
interacting with water and polycarbonate,
using a newly developed emulsion detector.
Total and partial charge-changing cross sections for the production
of B, Be, and Li fragments were measured and compared
with both previously published measurements, and model predictions.
This study is of importance for validating
and improving carbon ion therapy treatment planning systems,
and for estimating the radiological risks for personnel on space missions,
since carbon is a significant component of the Galactic Cosmic Rays.
\end{abstract}



\maketitle

%
%

\section{Introduction}

To understand the fragmentation of heavy ions in high energetic nucleus-nucleus
interactions is very important for many different applications,
e.g., heavy ion radiotherapy~\cite{therapy,therapy2}
and shielding against GCR (Galactic Cosmic Rays) and SPE 
(Solar Particle Events) during manned missions outside 
the earth magnetic field~\cite{space}.
For example, carbon ions of several hundred MeV/nucleon are currently used
for radiotherapy at several facilities in the world.
Radiotherapy with beams of heavy ions provides highly localized 
dose distributions at the end of the range (called Bragg peak),
and is expected to be biologically more efficient than conventional
photon radiation or proton beams.
However, even if the heavy ions have the advantage of giving a localized
dose distribution from the primary particles,
the projectile fragmentation products can cause
an undesired dose beyond the Bragg peak,
which cause an unwanted dose to the healthy tissues.
This effect can be estimated if the charge-changing cross sections of 
carbon ions in tissue equivalent materials, e.g., water, is known.
Knowledge of fragmentation of carbon ions is also important
when estimating the radiological risks in space,
since the carbon is one of
the most abundant heavy ion components in GCR, and its fragmentation products
can make a significant contribution to the dose given to the personnel
in the space vehicles.
However, the fragmentation of carbon ion is not yet fully understood.
Several measurements of the charge-changing 
cross sections of carbon in water and tissue materials have been performed~\cite{Sch,Gol}.
If compared to the predictions of different models~\cite{Sih,Wil,Tri,Brohm,Tsao,Sum},
significant discrepancies are observed.
Moreover, measurements of partial projectile fragmentation cross sections are scarce.
More experimental data are therefore needed to better understand
the fragmentation of carbon ion.

From the experimental point of view, emulsion is one of the most suitable devices
for detailed characterization of the fragmentation reactions,
since it has the capability of spatial resolution of micron-order
and excellent multi-particle separation in nearly 4$\pi$ solid angle.
Despite these advantages, this technology has not been applied in the past to
a high statistic experiment because the track recognition is quite time-consuming
due to the manual scanning of emulsion.
However, the recent success of developing an automatic
image processing system called Ultra Track Selector (UTS)~\cite{TS}
and an event reconstruction software package (NETSCAN)~\cite{NETSCAN}
improved the scanning procedure greatly.
Emulsion detectors are now used as core devices for large-scale 
elementary particle experiments~\cite{NETSCAN, chorus, OPERA}.
Taking advantage of this new emulsion technique,
we conducted an experiment to measure fragmentation of carbon ions at HIMAC
(Heavy Ion Medical Accelerator in Chiba) at NIRS 
(National Institute of Radiological Sciences), Chiba, Japan. 
In this experiment, we constructed an emulsion detector,
called Emulsion Cloud Chamber (ECC),
which has the ability of three dimensional track and 
vertex reconstruction and charge identification for each individual track. 

In this paper, we report on measurements of total and partial
charge-changing cross sections for incident $^{12}$C beams in water
and polycarbonate at energies ranging from 200 to 400 MeV/nucleon.
The results are discussed and compared with previous measurements and model predictions.

%

\section{Experiment}

\subsection{Emulsion Cloud Chamber}

\begin{figure}[t]
\begin{center}
\includegraphics[scale=.7]{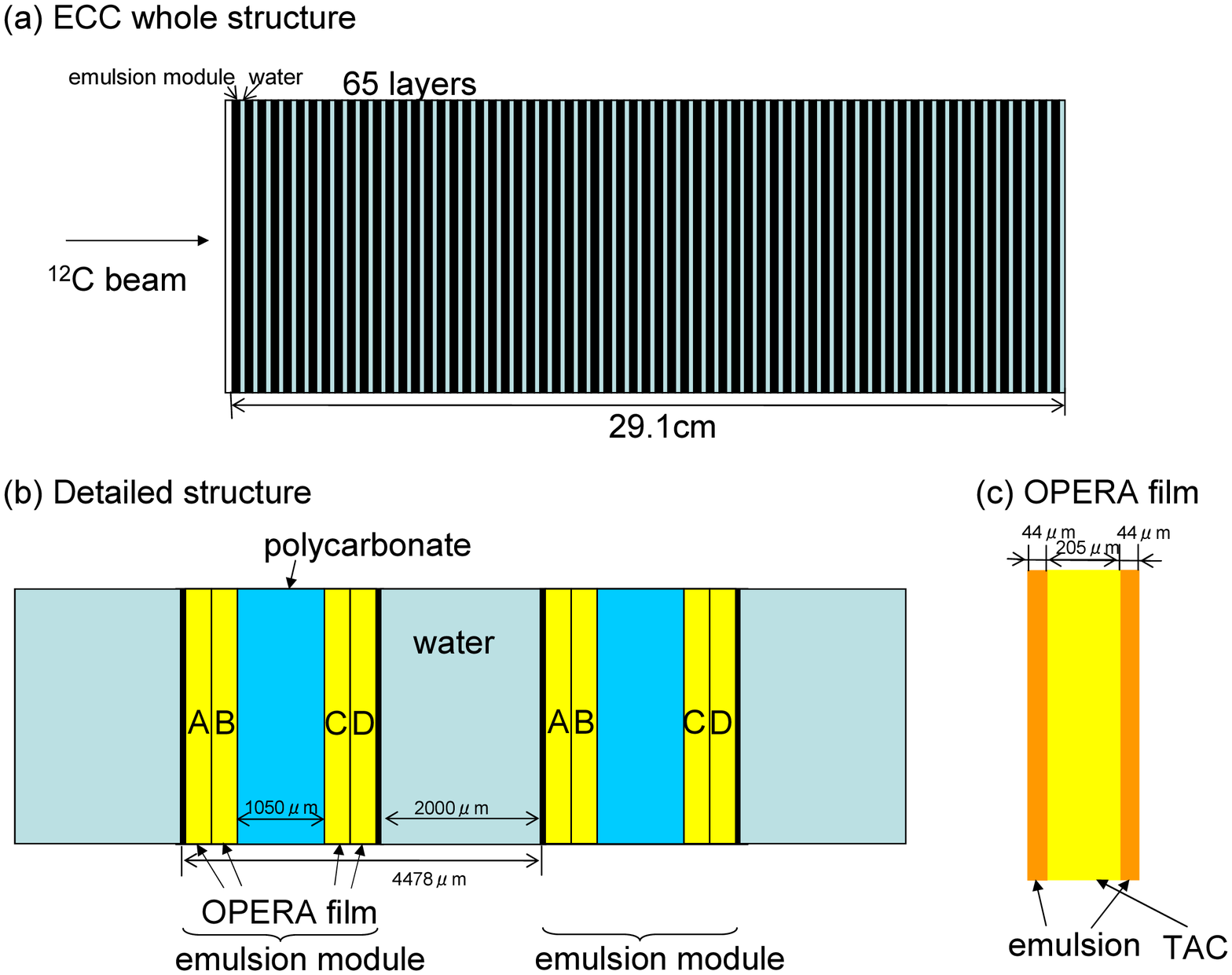}
\caption{\label{ecc}
(Color online) 
(a) A schematic view of the whole structure of the Emulsion Cloud Chamber (ECC).
It has 65~layers of emulsion modules and water targets.
(b) Detailed structure of the ECC.
An emulsion module consists of four emulsion sheets and a polycarbonate plate.
(c) Cross section of an OPERA film.
An OPERA film has emulsion layers coated on both sides of a TAC base. 
}
\end{center}
\end{figure}

The Emulsion Cloud Chamber (ECC) was designed so that it can store information
about all fragmentation reactions over the whole region where primary carbons
penetrate the chamber.
Schematic drawings of the ECC are shown in Fig.~\ref{ecc}.
The ECC has 65~layers\footnote{%
Consecutive numbers are assigned
to the layers along the beam direction.} 
of emulsion modules and acrylic frames,
and its total length is 29.1~cm long.
The acrylic frame has a thickness of 2~mm and its inside was filled with de-ionized water
at the exposure time.
The ECC was installed in a water tank which has a window of 1-mm-thick acrylic plate
on the side of beam injection.

Each emulsion module has four emulsion sheets reinforced
with a 1.05-mm-thick polycarbonate (C$_{16}$H$_{14}$O$_3$) plate whose measured density
is 1.16~g/cm$^{3}$.
A pair of emulsion sheets covers on both sides of the plate as shown
in Fig.~\ref{ecc}(b).\footnote{%
The labels ``A'' to ``D'' are assigned to emulsion
sheets in each module as shown in Fig.~\ref{ecc}(b).}
The polycarbonate plate keeps the flatness of the emulsion module,
though it was also used as a target material. Each emulsion module
is vacuum-packed with an aluminum-coated film for light and water shielding.
The OPERA film~\cite{operafilm} - a nuclear emulsion film developed
for the OPERA experiment~\cite{OPERA},
was used as the emulsion sheet.
An OPERA film measures 102~mm by 127~mm
and has 44-$\mu$m-thick emulsion layers coated on both sides of
a 205-$\mu$m-thick TAC (cellulose triacetate) base as shown in Fig.~\ref{ecc}(c).

\subsection{Beam exposure and development}
The beam exposure was performed in the SB2 beam line~\cite{secbeam} of
the HIMAC heavy ion synchrotron in December 2004. The primary $^{12}$C beam
with a kinetic energy of 400~MeV/nucleon was extracted to the beam line
and directly delivered to the experimental area.

The beam size and intensity were optimized so that the 
surface of ECC corresponding to the scanning area (12.5 cm$^{2}$)
was irradiated uniformly with a proper density.
Beam density is one of the most critical parameters for the exposure,
and exposures at excessive densities cause overlapping of particles,
which makes the track reconstruction difficult.
We controlled the beam density to be of the order of
10$^{3}$~particles/cm$^{2}$
which is the optimal density to observe fragmentation reactions
at energies above 200~MeV/nucleon.
We placed a 1-mm-thick plastic scintillator between the end of
the beam pipe and the ECC,
to monitor the number of injected charged particles.
The total count of beam tracks was 23874, and the beam density was measured
to be 1.6$\times$10$^{3}$~particles/cm$^{2}$.
The purity of $^{12}$C of the beam at the front of the ECC was 
estimated to be 99.5\% by taking into account the interactions of $^{12}$C
in the materials between the vacuum beam pipe and the ECC.

After the exposure, the emulsion sheets were shipped to the scanning facility
at Nagoya University.
Sheets-A and C were developed normally and sheets-B and D were processed
applying so-called the refreshing method~\cite{refresh}
before development.
Refreshing is a technology to erase minimum ionizing tracks
in emulsion by means of forced fading
under a high temperature and high humidity environment.
It desensitizes emulsion
and extends the dynamic range of response to highly charged particles.
In Ref.~\cite{refresh} we showed that charges of ion tracks 
can be identified up to Z~=~6 with the refreshing method.
Sheets-B and D were refreshed for three days
under the temperatures of 40~$^{\circ}$C and 38~$^{\circ}$C respectively
and the relative humidity of 98\%.

%

\section{Analysis}

\subsection{Emulsion read-out}
We used the UTS (Ultra Track Selector)~\cite{TS} to read out track images
in the emulsion films. UTS takes 16~layers of tomographic CCD images from
a 44-$\mu$m-thick emulsion layer of a film.
The imaging processor of UTS automatically finds track segments in an emulsion layer
as three dimensional vectors.
The UTS system has the field view of 150~$\mu$m by 120~$\mu$m and processes
three views per second.
Track segments with the tangents of the polar angles less than 0.5 were recorded,
and the detection efficiency of a track segment was 98\%.
After the scanning, two corresponding track segments
in the emulsion layers of both sides were connected across the TAC base,
a new track segment was thus created.
This is called a base track.
The positional and angular resolutions of a base track
were found to be 1~$\mu$m and 6~mrad, respectively.

UTS also records the grain density of a track segment,
which keeps information on the local energy deposited.
The grain density is digitized to pixel counts of CCD images whose pixel size
is 0.29~$\mu$m by 0.23~$\mu$m.
The pulse height is defined as the total pixel counts along each base track
and used to identify a track charge at the later stage.

\subsection{Track and vertex reconstruction}
The NETSCAN software framework~\cite{NETSCAN} was utilized for track
and vertex reconstruction.
NETSCAN reconstructs trajectories of charged particles by connecting base tracks
which lay within a certain distance.
At the tracking stage, only base tracks in sheets-A were used.
The probability of misconnection of two base tracks in adjacent modules
was expected to be less than 1\% for the primary beam tracks.
The base tracks in sheets-B and D, which were desensitized with the refreshing method,
were matched and associated with each reconstructed track. 

After the track reconstruction,
the three dimensional vertex reconstruction was performed.
Tracks starting inside the ECC and having a certain distance from a beam particle
were selected as secondaries.
Then, the vertex position was calculated searching for the position
at which multiple tracks converged.
We also imposed that an impact parameter of each secondary track should
be less than 20~$\mu$m,
which was determined to associate the secondary tracks having the energy
above 200~MeV/nucleon with a correct vertex
at more than 99\% efficiency
by considering the effect of multiple coulomb scattering. 
The position resolution of a vertex was found to be 50~$\mu$m
in the beam direction.

\subsection{Charge identification}

\begin{figure}[t]
\begin{center}
\includegraphics[scale=.8]{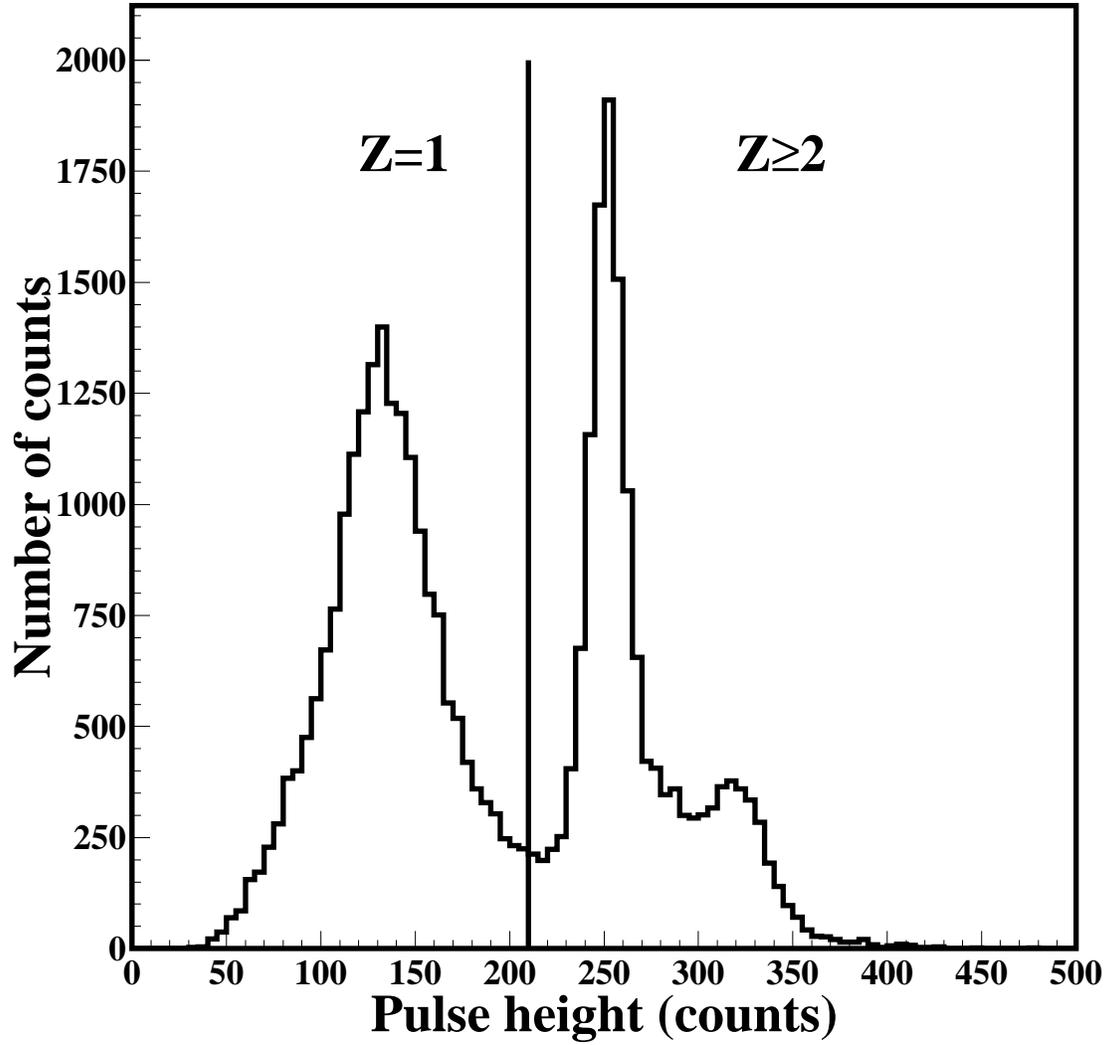}
\caption{\label{ph}
Distribution of the pulse heights in sheets-A for secondary tracks.
The tracks with pulse heights less than 210 
are discriminated as single charged tracks.
}
\end{center}
\end{figure}

\begin{figure}[t]
\begin{center}
\includegraphics[scale=.8]{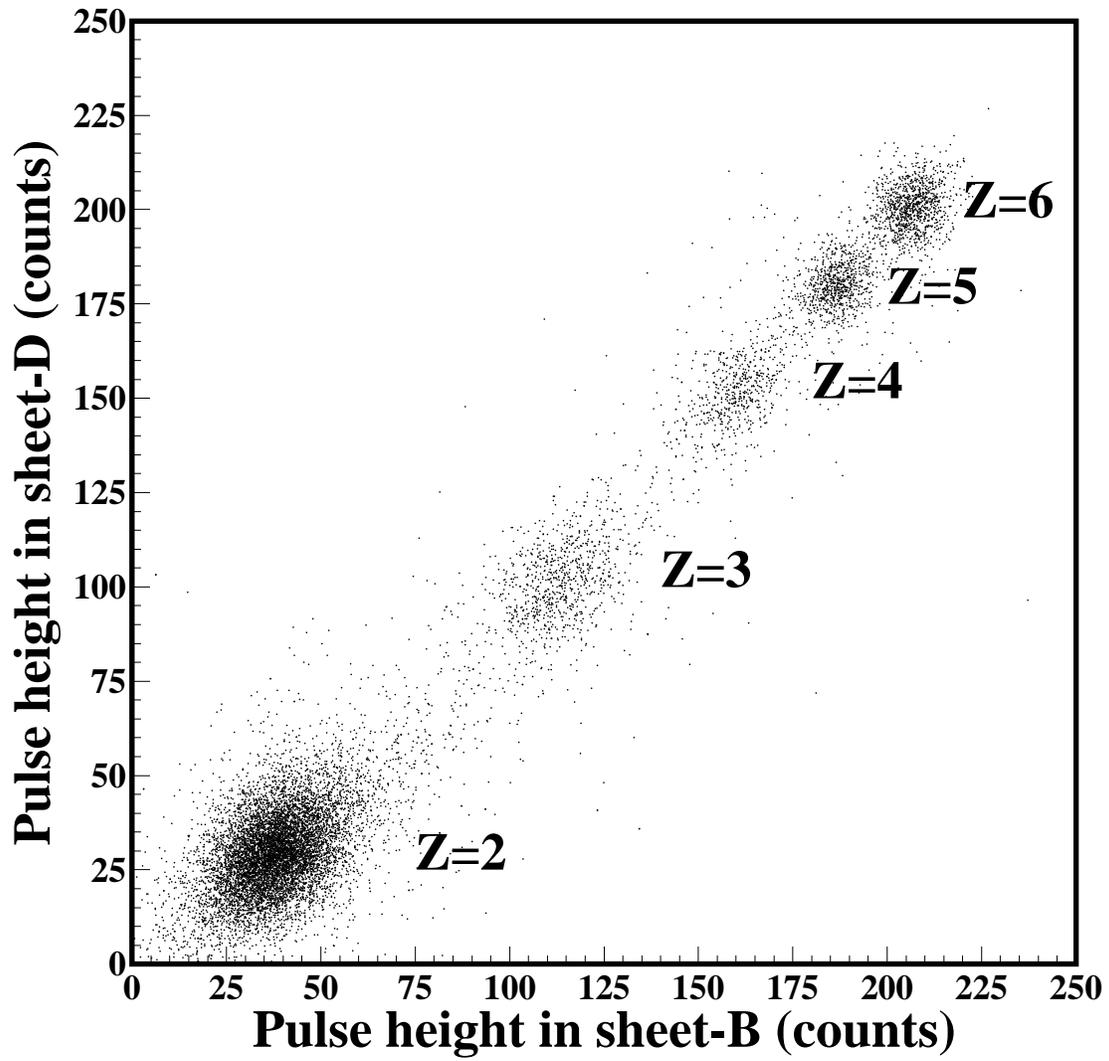}
\caption{\label{ph2}
A scatter plot between averaged pulse heights in sheets-B 
and sheets-D for secondary tracks.
One can see clear discrimination of track charges 
from Z = 2 to Z = 6.
}
\end{center}
\end{figure}

Pulse height information was used to identify the charge of each track.
Figure~\ref{ph} shows the pulse height distribution for secondary tracks,
where the pulse heights are averaged among the connected base tracks
in non-refreshing films of Sheets-A. As seen in the figure,
pulse heights for multiple charged tracks overlap with each other
and the track charges cannot be distinguished by this data alone.
We therefore determined the track charges in two steps.
At a first step, single charged tracks were discriminated from multiple charged tracks,
demanding that the averaged pulse heights are less than 210 as shown in Fig.~\ref{ph}.
We estimated that 6\% of single charged tracks were misidentified
as multiple charged tracks by this separation.
Then, as a second step, we used pulse height information of refreshed films.
The refreshing method desensitizes the emulsion films and decreases
the pulse heights depending on refresh conditions. Sheets-B and sheets-D
were refreshed in different conditions,
so that they would show different responses to ion charges.
Figure~\ref{ph2} presents a scatter plot between pulse heights averaged
in sheets-B and sheets-D for secondary tracks.
Clear discrimination of track charges from Z~=~2 to Z~=~6
can be obtained as seen in the figure.
Likelihood analysis
was applied for actual charge determination.
The likelihood function is defined as
\begin{displaymath}
\displaystyle
L(Z)  \equiv \prod_{i=B,D} P_i(Z,x),
\end{displaymath}
where $P_i(Z,x)$ is the probability function of a pulse height ($x$)
for a specific charge ($Z$) and a refresh condition ($i$).
We assumed that the pulse height distributions were Gaussian,
where the mean pulse height and its deviation for individual charge
were estimated by measured values for each refresh condition.
Finally, the track charges were determined by varying the $Z$
from two to six and finding the value which gave 
the maximum of the above likelihood function.
The separation between Z~=~5 and Z~=~6 is most crucial,
and the misidentification probability was 
estimated from the breadth of the pulse height distributions
and found to be 1.4\%.

\subsection{Counting of charge-changing interactions}
\begin{figure}[t]
\begin{center}
\includegraphics[scale=.9]{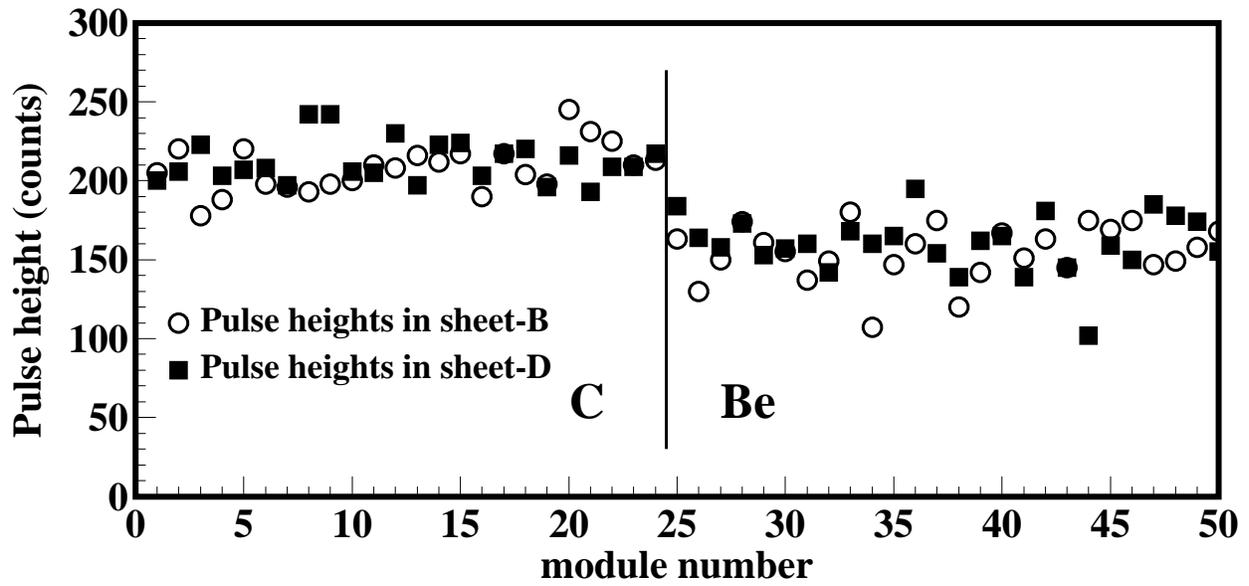}
\caption{\label{cchange}
An example of a charge-changing track.
Open circles (Black squares) represent averaged pulse heights in
sheets-B (sheets-D).
We traced pulse height variations along primary carbon tracks.
In this case, a carbon beam changes into a beryllium 
between 24th and 25th~modules.
}
\end{center}
\end{figure}

The charge-changing interactions were identified principally by finding vertices
whose primary tracks were observed as Z~=~6 and any of its secondary tracks were not Z~=~6.
To ensure beam tracks, primary tracks were also required to be connected
from the uppermost module.
However a certain number of charge-changing interactions
were missed using this identification condition.
For instance, if a primary carbon and a secondary track were reconstructed
as a single straight track, then the vertex would not be found.
The following case is conceivable; a primary carbon changes to a secondary fragment
with a kink angle too small to detect
and no other secondary tracks are found in the angular acceptance (tan$\theta < 0.5$).
In order to recover these kinds of events, pulse height variations were
traced along primary carbon tracks.
Figure~\ref{cchange} shows an example of this type of track
(we call it a ``charge-changing track'').
As seen in the figure, a charge-changing interaction can be detected
as a discrete change of pulse heights like a step function.
It turned out that 87\% of the interactions were detected as reconstructed vertices
and the rest were counted as charge-changing tracks.

\begin{figure}[t]
\begin{center}
\includegraphics[scale=.8]{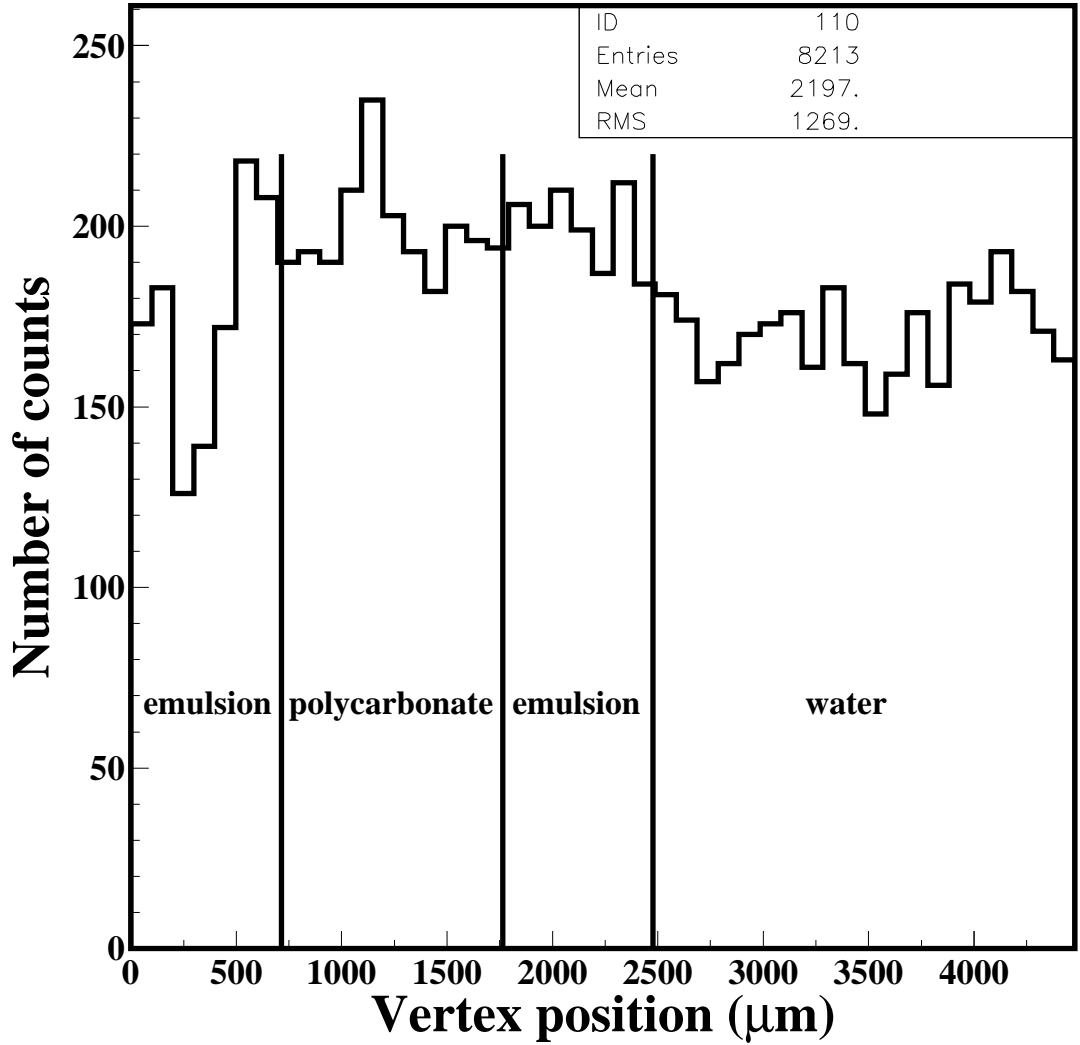}
\caption{\label{dz}
Distribution of local distance between the surface of each module and 
reconstructed vertices.
The target materials of interactions were identified 
according to the detector structure
(also see Fig.~\ref{ecc}(b)).
}
\end{center}
\end{figure}

For reconstructed vertices,
we can recognize carbon-water and carbon-polycarbonate interactions
by the vertex positions.
Figure~\ref{dz} shows the distribution of local distance
between the surface of each module and reconstructed vertices (also see Fig.~\ref{ecc}(b)).
The target materials of interactions were identified according to the detector structure
as shown in the figure.
For charge-changing tracks, we cannot directly determine the target material,
because we cannot get the exact vertex position.
To evaluate the number of interactions,
we did the following.
The number of interactions, $N_T(i)$,
in a target material of $T$ (water or polycarbonate) 
of $i$-th module can be obtained by
\begin{displaymath}
N_T(i) = N_T^{rec}(i) + R_T N^{cct}(i),
\end{displaymath}
where $N_T^{rec}$ denotes the number of reconstructed vertices
in a target ($T$) and $N^{cct}$ is the number of charge-changing tracks.
Here $R_T$ is introduced as the ratio of reactions in water or polycarbonate
to those in the whole region of the ECC.\footnote{In Fig.~\ref{dz},
one can see a deterioration in vertex reconstruction at the first emulsion layer.
This is due to interactions inside the emulsion layers which were used for tracking.
Actual counts of interactions in the first emulsion region were taken as those
in the second region.}
The ratio was given for each charge-changing;
$\Delta Z$~=~1, 2, 3, 4, and 5.
The counting inefficiency due to problems in reconstruction of secondary particles
was estimated to be less than 2\%. 

In the lower energy region,
the track and vertex reconstruction deteriorates due to larger scattering.
As effective events, we analyzed interactions in the upstream modules 
from 35th module,
which corresponds to the beam energy above 200~MeV/nucleon.
Finally, we obtained 8213 interactions in this region of the ECC.

\section{Results and Discussion}

\subsection{Total charge-changing cross section}
\begin{figure}[t]
\begin{center}
\includegraphics[scale=.8]{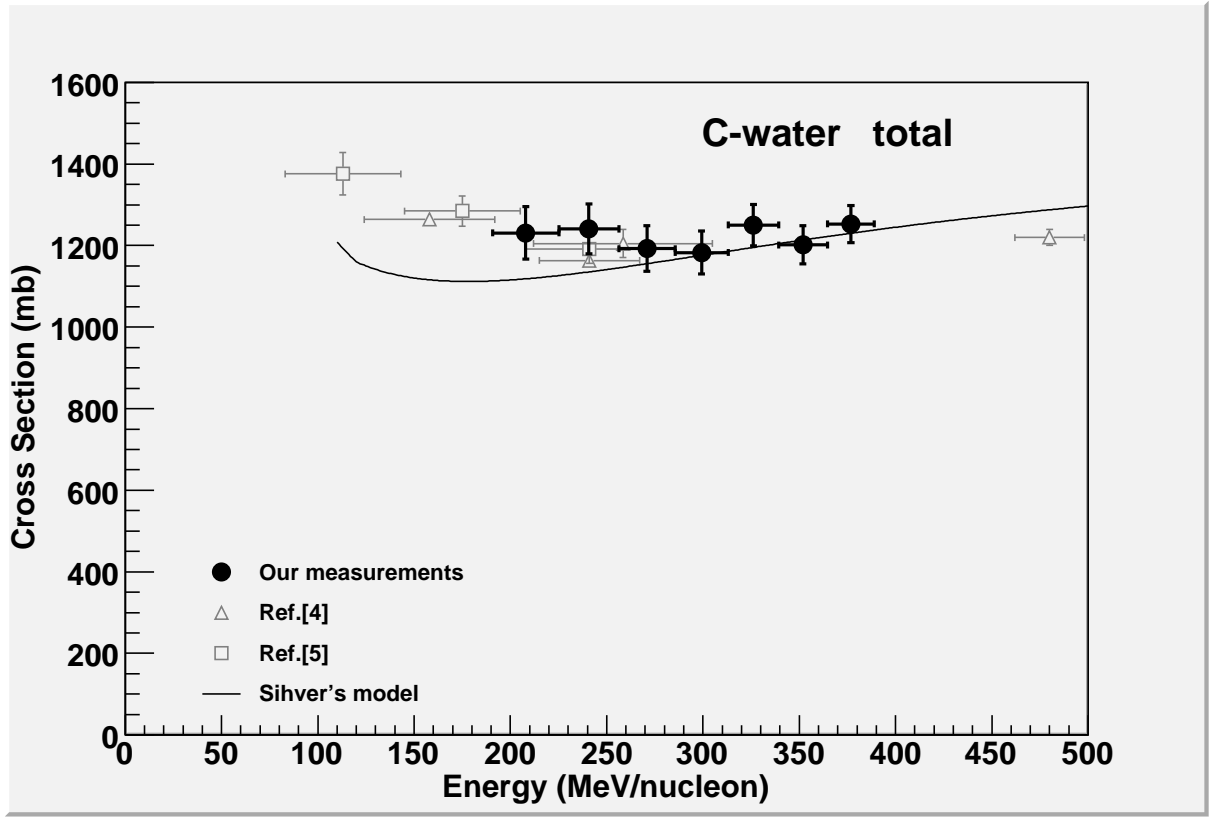}
\caption{\label{tcc}
Total charge-changing cross sections of carbon-water interactions
in comparison with previous measurements and a model calculation.
The vertical error bars indicate statistical errors only
in this and the followings figures.
}
\end{center}
\end{figure}

\begin{figure}[t]
\begin{center}
\includegraphics[scale=.8]{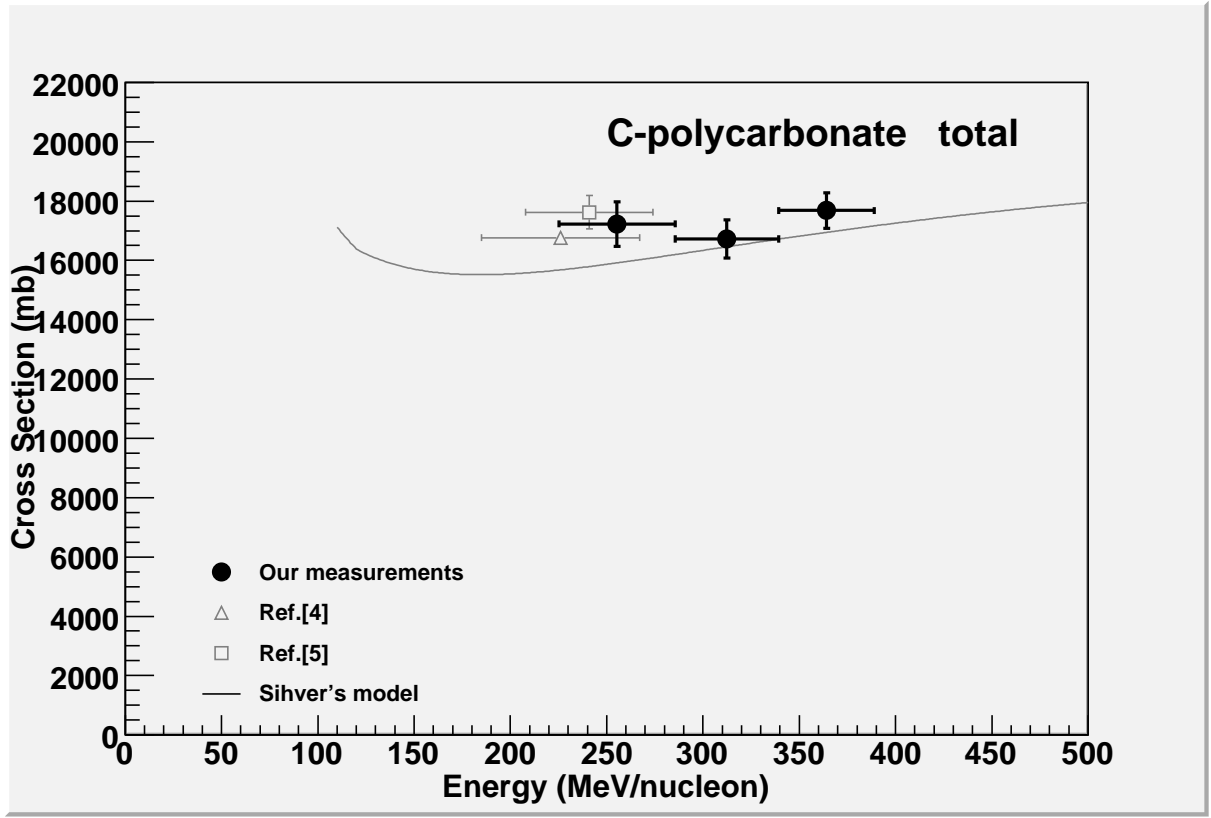}
\caption{\label{tccpc}
Total charge-changing cross sections of carbon-polycarbonate interactions
in comparison with previous measurements and a model calculation.
}
\end{center}
\end{figure}

The total charge-changing cross section ($\sigma_{tot}$)
for a target material $T$ in $i$-th module can 
be approximated by
\begin{displaymath}
\sigma_{tot}(i) = \frac{M}{N_A \rho t} 
\frac{N_T(i)}{N_C(i)},
\end{displaymath}
where $N_A$ is Avogadro's number,
$M$ is the molecular mass in atomic units of the target material,
$\rho$ is the target mass density, $t$ is the target thickness,
and $N_C(i)$ is the number of primary carbon tracks impinging on $i$-th module.
To reduce statistical errors, successive five (ten) modules are combined
for carbon-water (carbon-polycarbonate) interactions.
In Figs.~\ref{tcc} and ~\ref{tccpc} our measured total charge-changing cross sections,
from the interactions of carbon with water and polycarbonate,
are plotted together with measurements from Ref.~\cite{Sch} and Ref.~\cite{Gol}.
These cross sections and their statistical errors are also listed
in Tables~\ref{tab:twater} and \ref{tab:tpc}.
The cross sections of the carbon-polycarbonate (C-C$_{16}$H$_{14}$O$_3$) 
reactions in the references are obtained from the combination of cross sections
in different target materials; carbon-carbon (C-C), 
carbon-paraffin (C-CH$_{2}$), and carbon-water (C-H$_{2}$O) interactions. 
The beam energy at each module was estimated by energy loss calculation
using Geant4~\cite{Geant4}.
As can be seen, our measurements agree well with the previous measurements
at overlapping energies around 250 MeV/nucleon.

We have also compared our results with calculations based on the Sihver's model~\cite{Sih},
which are superimposed in Figs.~\ref{tcc} and \ref{tccpc}.
Sihver's model is a semi-empirical model which takes advantage of the experimentally 
verified weak factorization property~\cite{Olson}.
The model is used in the treatment planning systems for heavy ion radiotherapy
at several facilities in the world.
It is also used in models for calculations of Galactic Cosmic Rays, etc.
As can be seen in Figs.~\ref{tcc} and \ref{tccpc},
the total charge-changing cross sections
are quite consistent with the model predictions
at energies above 250 MeV/nucleon,
while the predictions differ from our results 
by about 10\% in the lower energy region.

The systematic error in our measurements consists of the following components:
\begin{enumerate}
\item The uncertainty in vertex positions affected the position cut
for selecting a target material, and resulted in an uncertainty of 3\%. 
\item The probability of charge misidentification of primary or secondary particles 
gave the uncertainty of 5\%.
\item The inefficiency of counting charge-changing reactions was taken as
the fractional systematic error of 2\%.
\end{enumerate}
The contamination of $^{10}$C and $^{11}$C in the $^{12}$C beam 
was expected to be 3\%.
Assuming that the differences of the cross sections between these isotopes 
are small,
the contribution to the systematic error is considered to be 
negligibly small~\cite{Gol}.
By summing up all these uncertainties,
we get an estimated total systematic error of 6\%.

\subsection{Partial charge-changing cross section }

\begin{figure}[t]
\begin{center}
\includegraphics[scale=.8]{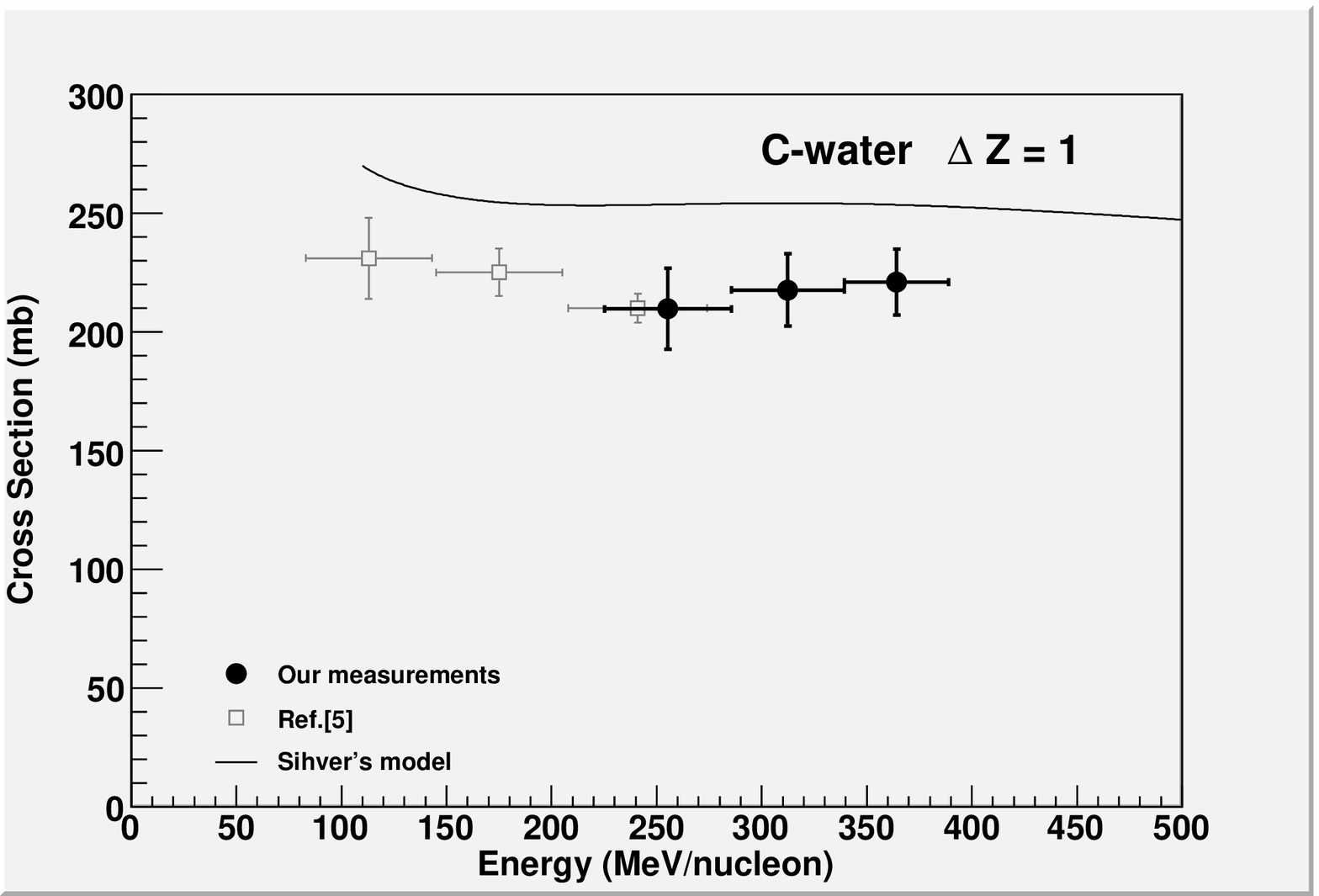}
\caption{\label{pcc5}
Partial charge-changing cross sections 
of carbon-water interactions of $\Delta Z$~=~1 
in comparison with previous measurements and a model calculation.
}
\end{center}
\end{figure}

\begin{figure}[t]
\begin{center}
\includegraphics[scale=.8]{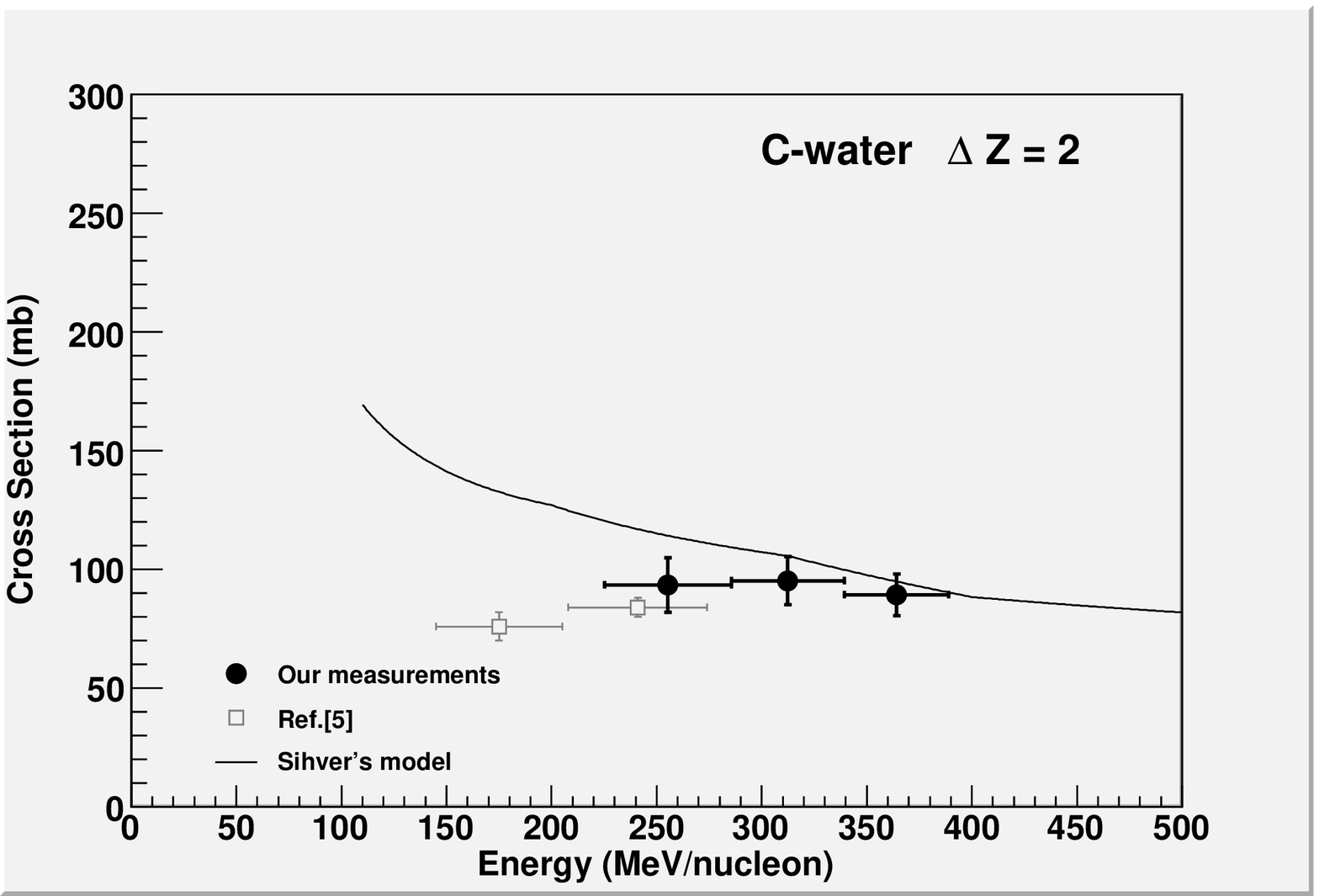}
\caption{\label{pcc4}
Partial charge-changing cross sections 
of carbon-water interactions of $\Delta Z$~=~2 
in comparison with previous measurements and a model calculation.
}
\end{center}
\end{figure}

\begin{figure}[t]
\begin{center}
\includegraphics[scale=.8]{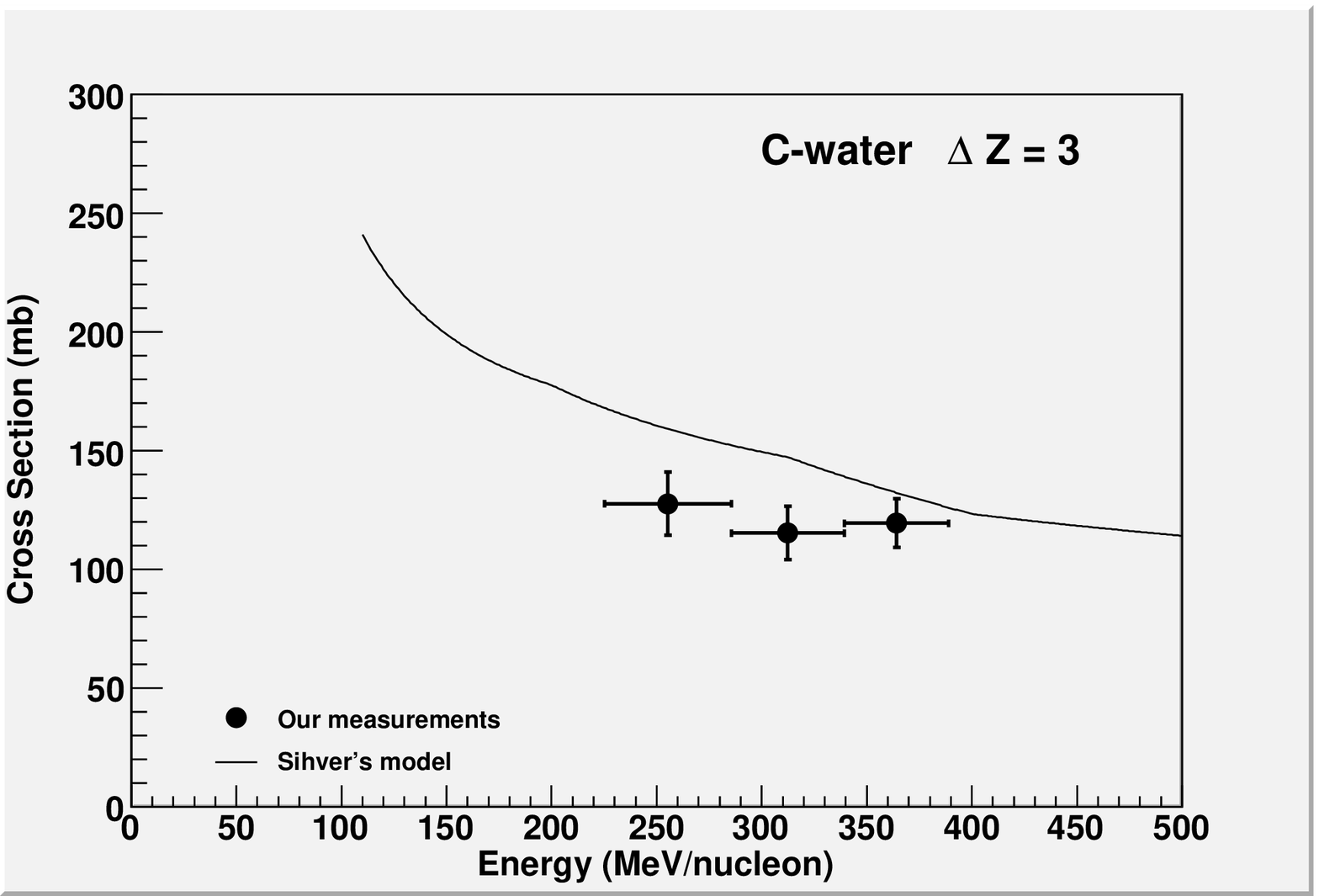}
\caption{\label{pcc3}
Partial charge-changing cross sections 
of carbon-water interactions of $\Delta Z$~=~3
in comparison with a model calculation.
}
\end{center}
\end{figure}

\begin{figure}[t]
\begin{center}
\includegraphics[scale=.8]{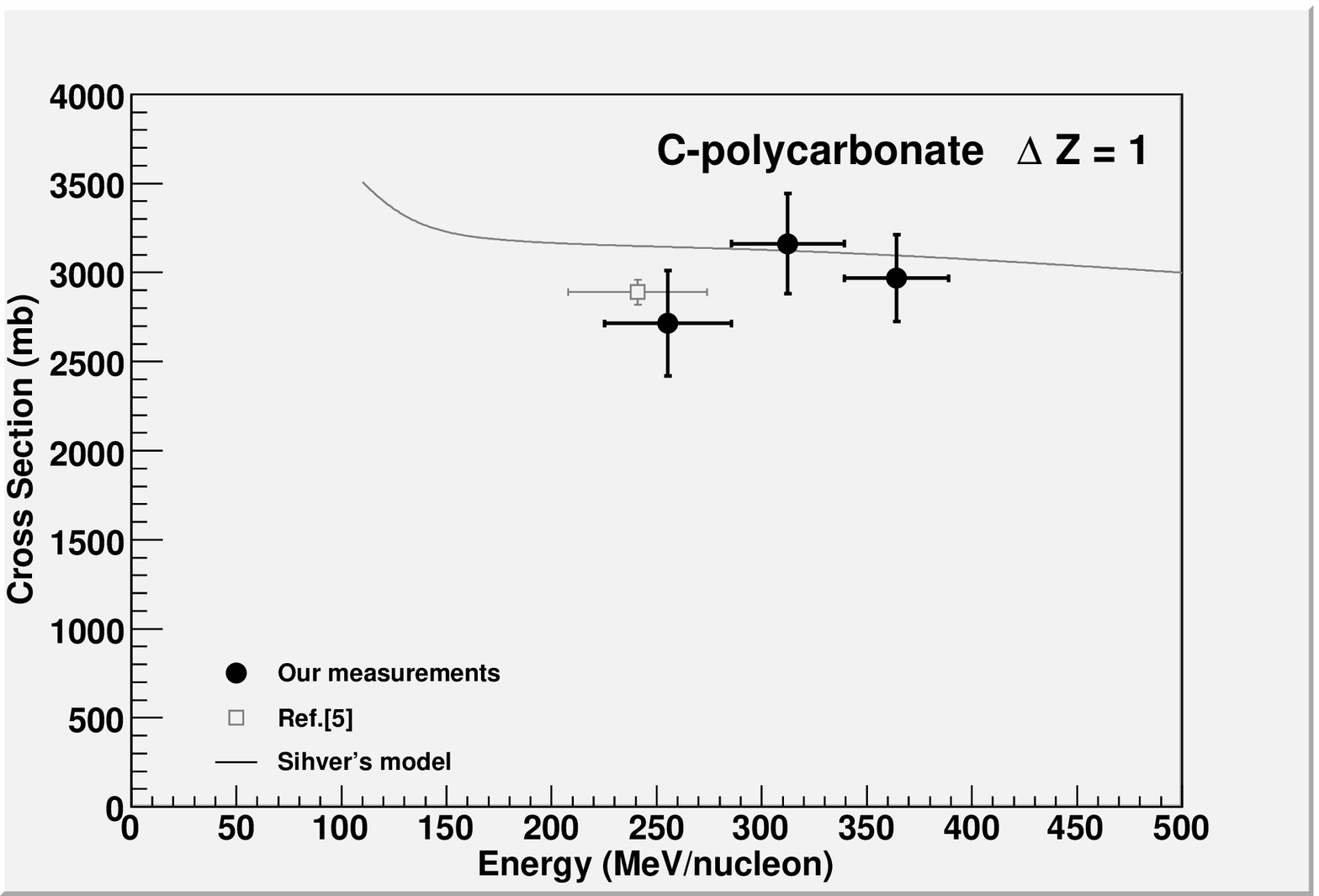}
\caption{\label{pcc5pc}
Partial charge-changing cross sections 
of carbon-polycarbonate interactions of $\Delta Z$~=~1 
in comparison with previous measurements and a model calculation.
}
\end{center}
\end{figure}

\begin{figure}[t]
\begin{center}
\includegraphics[scale=.8]{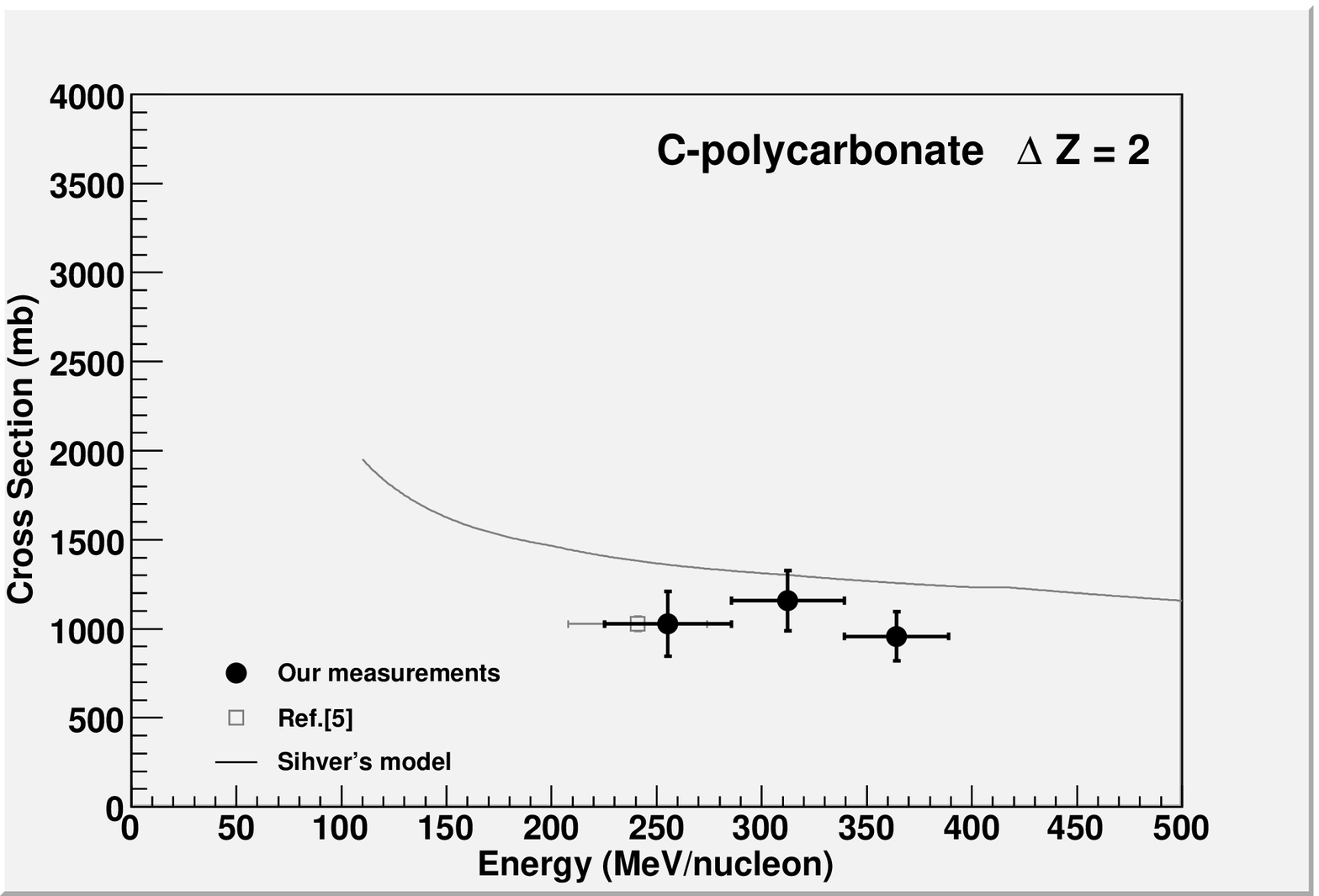}
\caption{\label{pcc4pc}
Partial charge-changing cross sections 
of carbon-polycarbonate interactions of $\Delta Z$~=~2 
in comparison with previous measurements and a model calculation.
}
\end{center}
\end{figure}

\begin{figure}[t]
\begin{center}
\includegraphics[scale=.8]{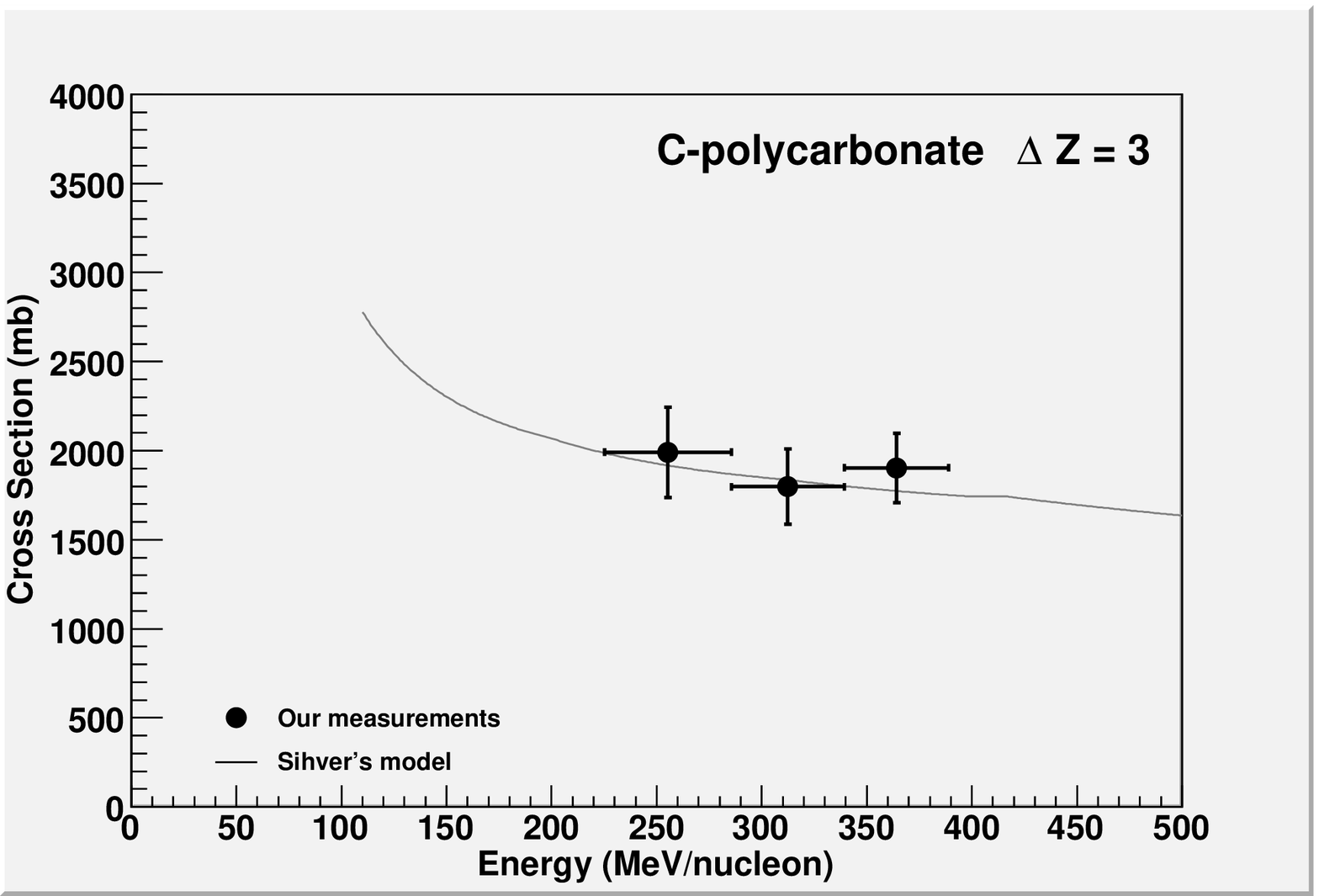}
\caption{\label{pcc3pc}
Partial charge-changing cross sections 
of carbon-polycarbonate interactions of $\Delta Z$~=~3
in comparison with a model calculation.
}
\end{center}
\end{figure}

The partial projectile charge-changing cross section ($\sigma_{\Delta Z}$)
is defined as the cross section producing a projectile-like fragment
with a specific charge difference
$\Delta Z$ = 6 - $Z_{\mathrm{fragment}}$, ranging from $\Delta Z$ = 1 for B to $\Delta Z$ = 5 for protons.
Figures~\ref{pcc5}-\ref{pcc3} and ~\ref{pcc5pc}-\ref{pcc3pc}
show our measurement results for carbon reactions of
$\Delta Z$ = 1, 2, and 3 in targets of water and polycarbonate, respectively.
In Figs.~\ref{pcc5},~\ref{pcc4},~\ref{pcc5pc}, and~\ref{pcc4pc},
our measured partial charge-changing cross sections are
also compared with the measurements published in Ref.~\cite{Gol}.
All cross sections are also listed in Tables~\ref{tab:pwater} and \ref{tab:ppc}.
As can be seen from these figures, our measurements agree well with previously
performed measurements at overlapping energies around 250 MeV/nucleon.

We have also compared with calculations based on the Sihver's model~\cite{Sih},
which are superimposed in Figs.~\ref{pcc5}-\ref{pcc3pc}.\footnote{The values of Sihver's model
were obtained by summing up for all
detectable isotopes with A = 11, 10, 8 for $\Delta Z$ = 1,
A = 10, 9, 7 for $\Delta Z$ = 2, and A = 7, 6 for $\Delta Z$ = 3.}
The model predictions for the partial cross sections show stronger 
energy dependence than our measurements, 
and we found significant discrepancies,
especially in the lower energy region.
The deviations between the measured and the calculated cross sections
for the fragments with $\Delta Z$ = 1, 2, and 3 at the projectile energy 255 MeV/nucleon,
are 21\%, 23\%, and 24\% for the carbon-water interactions,
and 16\%, 33\%, and 4\% for carbon-polycarbonate interactions, respectively.
The total systematic error was estimated to be 6\%,
in the same way for the total cross sections. 

We do not present the results of the partial charge-changing cross sections
for $\Delta Z$ = 4 and 5 i.e., the production of $\alpha$ and p,
since these light particles have energy and
angular distributions beyond the detector acceptance;
and contamination of light target fragments can not be neglected.

\section{Conclusions}

We have measured charged nuclear fragments produced by 200 - 400 MeV/nucleon carbon ions,
interacting with water and polycarbonate,
using a newly developed emulsion technique.
Both total and partial charge-changing cross sections were measured,
and the results were compared with previous measurements and model predictions.
The measured total charge-changing cross sections showed good agreement
with both the reference measurements and simulations using a model
developed by Sihver et al.~\cite{Sih}.
In the case of the partial cross sections
Sihver's model showed stronger energy dependence
than our measurements, and deviated from our measurements in the lower energy region.
This proves that measurements performed by our ECC can help improve
and validate theoretical fragmentation models.
Since 200 - 400 MeV/nucleon
is the energy region where carbon ion therapy is performed at several facilities,
our data is of importance for validating and improving the treatment planning systems.
The results are also of importance when estimating the radiological risks
for personnel on space missions, since carbon is a significant component of the GCR.

\section*{ACKNOWLEDGMENTS}
This research was performed as an Research Project with Heavy Ions at NIRS-HIMAC.
Also this work was supported in part by CREST of Japan Science and Technology Agency (JST).
We would like to thank B.~Lundberg for carefully reading this paper.
\newpage

%
%


\newpage
%
%

\nopagebreak[4]
\begin{table*}[tbp]
\caption{Comparison of total charge-changing cross sections 
of $^{12}$C beams in water target. }
\label{tab:twater}
\begin{center}
\begin{tabular}{cccccccc} \hline \hline
\multicolumn{2}{c}{Our measurements} &  &
\multicolumn{2}{c} {Ref.~\cite{Sch}} &  &
\multicolumn{2}{c}{Ref.~\cite{Gol}} \\ 
\cline{1-2} \cline{4-5} \cline{7-8}
E (MeV/nucleon) & $\sigma_{tot}$ (mb) & &
E (MeV/nucleon) & $\sigma_{tot}$ (mb) & &
E (MeV/nucleon) & $\sigma_{tot}$ (mb)  \\ \hline
377 $\pm$ 12 & 1253 $\pm$ 45  & & & & & & \\
352 $\pm$ 13 & 1202 $\pm$ 47  & & & & & & \\
326 $\pm$ 13 & 1250 $\pm$ 51  & & & & & & \\
299 $\pm$ 14 & 1183 $\pm$ 52  & & & & & & \\
271 $\pm$ 15 & 1193 $\pm$ 56  & & 259 $\pm$ 47 & 1205 $\pm$ 34  & & & \\
241 $\pm$ 16 & 1241 $\pm$ 61  & & 241 $\pm$ 26 & 1163 $\pm$ 13  
                              & & 241 $\pm$ 33 & 1191 $\pm$ 35 \\
208 $\pm$ 17 & 1231 $\pm$ 64  & & & & & & \\
\hline \hline
\end{tabular}
\end{center}
\end{table*}

\nopagebreak[4]
\begin{table*}[tbp]
\caption{Comparison of total charge-changing cross sections of 
$^{12}$C beams in polycarbonate (C$_{16}$H$_{14}$O$_3$) target. 
The values of Ref.~\cite{Sch} and Ref.~\cite{Gol} are obtained from 
the combination of cross sections in different target materials
(C, CH$_2$, and H$_2$O).}
\label{tab:tpc}
\begin{center}
\begin{tabular}{cccccccc} \hline \hline
\multicolumn{2}{c}{Our measurements} & &
\multicolumn{2}{c} {Ref.~\cite{Sch}} &  &
\multicolumn{2}{c}{Ref.~\cite{Gol}} \\ 
\cline{1-2} \cline{4-5} \cline{7-8}
E (MeV/nucleon) & $\sigma_{tot}$ (mb) & &
E (MeV/nucleon) & $\sigma_{tot}$ (mb) & &
E (MeV/nucleon) & $\sigma_{tot}$ (mb) \\ \hline
364 $\pm$ 25 & 17682 $\pm$ 595 & & & & & & \\
312 $\pm$ 27 & 16723 $\pm$ 648 & & & & & & \\
255 $\pm$ 30 & 17225 $\pm$ 747 & & 226 $\pm$ 41 & 16765  $\pm$ 134
                               & & 241 $\pm$ 33 & 17625 $\pm$ 562   \\
\hline \hline
\end{tabular}
\end{center}
\end{table*}

\nopagebreak[4]
\begin{table*}[tbp]
\caption{Comparison of partial charge-changing cross sections 
of $^{12}$C beams in water target. }
\label{tab:pwater}
\begin{center}
\begin{tabular}{cccccc} \hline \hline
& \multicolumn{2}{c}{Our measurements} & &
\multicolumn{2}{c}{Ref.~\cite{Gol}} \\ 
\cline{2-3} \cline{5-6}
$\Delta Z$& E (MeV/nucleon) & $\sigma_{\Delta Z}$ (mb) & &
            E (MeV/nucleon) & $\sigma_{\Delta Z}$ (mb) \\ \hline
1 & 364 $\pm$ 25 & 221 $\pm$  14 & & & \\
  & 312 $\pm$ 27 & 218 $\pm$  15 & & & \\
  & 255 $\pm$ 30 & 210 $\pm$  17 & & 241 $\pm$ 33 & 210 $\pm$ 6 \\
2 & 364 $\pm$ 25 & 89 $\pm$  9  & & & \\
  & 312 $\pm$ 27 & 95 $\pm$  10 & & & \\
  & 255 $\pm$ 30 & 93 $\pm$  11 & & 241 $\pm$ 33 & 84 $\pm$ 4 \\
3 & 364 $\pm$ 25 & 120 $\pm$  10 & & & \\
  & 312 $\pm$ 27 & 115 $\pm$  11 & & & \\
  & 255 $\pm$ 30 & 128 $\pm$  13 & & & \\
\hline \hline
\end{tabular}
\end{center}
\end{table*}

\nopagebreak[4]
\begin{table*}[tbp]
\caption{Comparison of partial charge-changing cross sections of 
$^{12}$C beams in polycarbonate (C$_{16}$H$_{14}$O$_3$) target. 
The values of Ref.~\cite{Gol} are obtained from 
the combination of cross sections in different target materials
(C, CH$_2$, and H$_2$O).}
\label{tab:ppc}
\begin{center}
\begin{tabular}{cccccc} \hline \hline
& \multicolumn{2}{c}{Our measurements} & &
\multicolumn{2}{c}{Ref.~\cite{Gol}} \\ 
\cline{2-3} \cline{5-6}
$\Delta Z$& E (MeV/nucleon) & $\sigma_{\Delta Z}$ (mb) & &
            E (MeV/nucleon) & $\sigma_{\Delta Z}$ (mb) \\ \hline
1 & 364 $\pm$ 25 & 2968 $\pm$  243  & & & \\
  & 312 $\pm$ 27 & 3162 $\pm$  281 & & & \\
  & 255 $\pm$ 30 & 2714 $\pm$  296 & & 241 $\pm$ 33 & 2890 $\pm$ 70 \\
2 & 364 $\pm$ 25 &  957 $\pm$  138 & & & \\
  & 312 $\pm$ 27 & 1158 $\pm$  170 & & & \\
  & 255 $\pm$ 30 & 1027 $\pm$  182 & & 241 $\pm$ 33 & 1028 $\pm$ 43 \\
3 & 364 $\pm$ 25 & 1903 $\pm$  195 & & & \\
  & 312 $\pm$ 27 & 1798 $\pm$  212 & & & \\
  & 255 $\pm$ 30 & 1992 $\pm$  253 & & & \\
\hline \hline
\end{tabular}
\end{center}
\end{table*}

\newpage
%
%

\end{document}